\documentclass[floatfix,twocolumn,aps,prd,showpacs,amsmath,amssymb]{revtex4}
\usepackage{epsfig}

\usepackage[latin1]{inputenc}
\begin{document}
\title{Excited state contribution to the Casimir-Polder force at finite temperature}

\author{T.N.C. Mendes and C. Farina}
\affiliation {Instituto de F\'{\i}sica,
UFRJ, Caixa Postal 68528,
21945-970 Rio de Janeiro RJ, Brazil}

\date{\today}
\begin{abstract}
Using the master equation we calculate the contribution of the excited state of a two-level atom to its interacting
potential with a perfectly conducting wall at finite temperature. For low temperature,
$\hbar \omega_0/k_B T = k_0 \lambda_T\gg 1$, where $\omega_0 = k_0 c$ is the transition frequency of the atom and
$\lambda_T$ is the thermal wavelength, we show that this contribution is very small $\left(\propto e^{-k_0\lambda_T}\right)$. In the opposite limit $(k_0\lambda_T \ll 1)$, however, we show that the expression for the interacting potential, for all relevant distance regimes, becomes exactly the same as that for very short distances $(k_0 z \ll 1)$ and with the field in the vacuum state.
\end{abstract}

\maketitle



\hyphenation{tem-pe-ra-tu-re}
\hyphenation{gi-ven}
\hyphenation{re-le-vant}
\hyphenation{ea-si-ly}
\hyphenation{stu-di-ed}
\hyphenation{star-ting}
\hyphenation{le-vel}
\hyphenation{slabs}
\hyphenation{re-fe-ren-ce}
\hyphenation{in-te-rac-ti-on}
\hyphenation{dis-cre-pan-cy}
\hyphenation{cons-tant}
\hyphenation{fi-gu-re}
\hyphenation{de-ve-lo-ped}
\hyphenation{se-ve-ral}
\hyphenation{Waals}

%

In 1948 Casimir and Polder \cite{CasiPolder} considered for the first time retardation effects on the dispersive van der Waals forces between two atoms and between an atom and a perfectly conducting wall. Since then, these forces are called Casimir-Polder forces (CP forces), and this subject have been explored exaustively in the literature. Good reviews have been written on dispersive van der Waals interaction \cite{Milonni,Langbein} and many elaborated papers concerning level-shifts near surfaces have appeared, as for instance, \cite{Meschede}, to mention just a few. It is worth mentioning that CP forces have been observed experimentally \cite{HindsetAl93}.

Recently, the influence of real conditions on the CP interaction has been considered \cite{Mostepanenko2004}. Further, higher multipole corrections \cite{Salam}, lateral Casimir forces \cite{PAm}, the influence of the CP interaction on Bose-Einstein condensates \cite{Vuletic,Antezza} and applications to nanotubes \cite{Nanotubo} are some of the many branches of great activity on this subject nowadays.

In a previous paper \cite{TarciroJPA2006} we analysed the interaction of an atom with a perfectly conducting wall starting from the general expressions for the energy level shifts of a small system interacting with a large one  considered as a reservoir. In that work we studied the vacuum and thermal contributions separately and considered only the level shift of the groundstate of the atom. In this letter we shall generalize our previous result taking into account the excited state
contribution. As far we know, our final result has never appeared in the literature. We shall follow the same procedure as that presented in reference \cite{TarciroJPB2006}.

Adopting  the dipole approximation the coupling between the radiation field and the atom is given by
$V({\bf x},t) = -{\bf d}\left( t\right) \cdot {\bf E}\left( {\bf x}, t\right)$, where ${\bf d}\left( t\right)$ is the dipole moment of the atom induced by the electric field ${\bf E}\left({\bf x},t\right) = \sum_{{\bf k}\lambda}{\bf F}_{{\bf k}\lambda}\left({\bf x},t\right)a_{{\bf k}\lambda}^{\dag}+h.c.$. In this expression, the field mode ${\bf F}_{{\bf k}\lambda}\left({\bf x},t\right)$ is a function that takes into account the contributions of sources and boundary conditions imposed to the field and $a_{{\bf k}\lambda}^{\dag}$ is the creation operator of a photon with wave-vector ${\bf k}$ and polarization $\lambda$ which satisfies the commutation relations $\left[a_{{\bf k}\lambda},a_{{\bf k}^{\prime}\lambda^{\prime}}^{\dag}\right]=\delta_{{\bf k}{\bf k}^{\prime}}\delta_{\lambda\lambda^{\prime}}$. Let $\vert g\rangle$ and $\vert e\rangle$ the ground and the excited states of the atom with unperturbed energies $E_g$ and $E_e$ respectively. Then, the level shifts are given by:
\begin{eqnarray}
\label{dErr(fr)}
\!\!\!\!\!\!\! &\delta&\!\!\!\!\!\!\! E_g^{rr}=\delta E_e^{rr}\;\;\; ; \;\;\;\; \delta E_g^{fr}=-\delta E_e^{fr}
\\
%
\label{Level_rr}
\!\!\!\!\!\!\! &\delta&\!\!\!\!\!\!\! E_g^{rr}= -{1\over 2}\sum_{{\bf k}\lambda}\alpha_{-}(k)\vert {\bf F}_{{\bf k}\lambda}\left({\bf x},t\right)\vert^2 
\\
%
\label{Level_fr}
\!\!\!\!\!\!\!&\delta&\!\!\!\!\!\!\! E_g^{fr}= -\sum_{{\bf k}\lambda}\alpha_{+}(k)\vert {\bf F}_{{\bf k}\lambda}\left({\bf x},t\right)\vert^2\!\left(\langle n_{{\bf k}\lambda}\rangle+{1\over 2}\right)
\\
%
\label{alpha}
\!\!\!\!\!\!\!&\alpha&\!\!\!\!\!\!\!_{\mp}\left( k\right)=\frac{\alpha_0 k_0}{2}\left(  {\mathcal P}\frac{1}{k + k_0} \pm {\mathcal P}\frac{1}{k - k_0}\right) \, ,
\end{eqnarray}
where $\delta E_{g(e)}=\delta E_{g(e)}^{fr}+\delta E_{g(e)}^{rr}$ is the level shift of the state $\vert g\rangle \left(\vert e\rangle\right)$, $\alpha_0=2\vert {\mathbf d}_{eg}\vert^2/3\hbar\omega_0$ is the static polariza\-bility, $\omega_0=k_0 c=\left(E_e-E_g\right)/\hbar$ is the transition frequency of the atom, $\mathcal P$ is the principal Cauchy value, $k=\vert {\bf k}\vert$ and $\langle n_{{\bf k}\lambda}\rangle$ is the statistical average number of photons in a given mode. Equation (\ref{Level_rr}) gives the {\it reservoir reaction} contribution and has the same value for both $\vert g\rangle$ and $\vert e\rangle$ states. Equation (\ref{Level_fr}), which gives the {\it fluctuations of reservoir} contribution, however, has opposite  signs for $\vert g\rangle$ and $\vert e\rangle$. Since this term is the only one that carries the information about the field state, we expect that the computation of the $\vert e\rangle$-contribution to the interacting potential will make it weaker.

In order to weight up the $\vert e\rangle$-contribution, let $p$ be the probability of the atom to be in the $\vert g\rangle$ state. Then $1-p$ will be the probability to be in the $\vert e\rangle$ state. Hence, the total average level shift of the atom is: $\delta E=\left(1-2 p\right) \delta E_g+2 p  \delta E_g^{rr}$. For the specific case of a two-level atom in a thermal bath at temperatute $T$ and separated by a distance $z$ of a perfectly conducting wall, last equation leads to:
\begin{eqnarray}
\label{weigh_Int}
V\left(z,T\right)\!\!\!\!&=&\!\!\!\! \tanh\! \left(\frac{1}{2} k_0\lambda_T\!\right)\! V_g\left(z,T\right)+{2 V_0^{rr}\left(z\right)\over e^{k_0\lambda_T}+1}\, ,\;\;\;\;\;\;
\\
%
\label{Vg}
V_g\left(z, T\right)\!\!\!\!&=&\!\!\!\! V^{fr}_g\left(z, T\right) + V^{rr}_0\left(z\right)\, ,
\nonumber \\
%
\label{Vgfr}
V^{fr}_g\!\left(z, T\right)\!\!\!\!&=&\!\!\!\! {\hbar c\over \pi}\!\int_0^{\infty}\!\! k^3\alpha_+\!\left(k\right)G\!\left(2k z\right)\coth\! \left(\!{1\over 2}k\lambda_T\!\right)dk\, ,
\nonumber \\
%
\label{Vgrr}
V^{rr}_0\left(z\right)\!\!\!\!&=&\!\!\!\! {\hbar c\over \pi}\!\int_0^{\infty}\!\! k^3\alpha_-\!\left(k\right)G\left(2k z\right) dk\, ,
\nonumber \\
%
\label{G}
G\left(x\right)\!\!\!\!&=&\!\!\!\! {\sin x\over x} + 2 {\cos x\over x^2} - 2 {\sin x\over x^3}\, ,
\nonumber
\end{eqnarray}
where $\lambda_T = \hbar c/k_B T$ is the thermal length, $k_B$ is the Boltzmann constant, $V_g\left(z,T\right)$ is the  interaction due to the ground-state contribution only, $V^{fr}_g\left(z,T\right)$ is the ``fluctuation-reservoir" contribution and $V_0^{rr}\left(z\right)$ is the ``reservoir-reaction" contribution \cite{TarciroJPA2006}. Recall that $\lambda_T$ defines a length scale beyond which the thermal contribution to the interaction is dominant relative to the vacuum contribution. Equation (\ref{weigh_Int}) is the main result of this letter, from which we shall derive some important particular cases.

Looking at distances smaller than $\lambda_T$, in the low temperature limit ($k_0\lambda_T \gg 1$), one may write (\ref{weigh_Int}) as
$$
V^{(\ell )}\left(z,T\right)\simeq V_g\left(z,T\right)-2 e^{-k_0\lambda_T}\left[V_g\left(z,T\right)-V_0^{rr}\left(z\right)\right]
$$
which implies
\begin{equation}\label{SL}
V^{(\ell)}_L\left(z,T\right) = -{\hbar \omega_0\over 8}{\alpha_0 \over z^3}+e^{-k_0\lambda_T}\mathcal O\left(z^{-2}\right)\, ,
\end{equation}
valid for $k_0 z\ll 1$ and
\begin{equation}\label{LL}
V^{(\ell)}_{CP}\left(z,T\right)  =
-{3\hbar c\over 8\pi}{\alpha_0 \over z^4}+e^{-k_0\lambda_T}\mathcal O\left(z^{-1}\right)\, ,
\end{equation}
valid for $1\ll k_0 z\ll k_0\lambda_T$. From  now on, $V^{(\ell)}_L$ and $V^{(\ell)}_{CP}$ will be referred to as London and Casimir-Polder regimes for low temperature respectively. In the last two equations, the $\vert e\rangle$-contribution is very small compared to the $\vert g\rangle$-contribution. This is expected since for $k_B T \ll \hbar \omega_0$ the probability of finding the system in its ground-state is much larger than to find it in its excited-state.

For distances larger than the thermal length, $z>\lambda_T$, and for any value of temperature, equation (\ref{weigh_Int}) may be written as:
\begin{equation}
\label{V_L}
V_{z>\lambda_T}\left(z,T\right) = -{\hbar \omega_0\over 8}{\alpha_0\over z^3}f\left(\theta\right)
\end{equation}
where $f\left(\theta\right)=\theta \tanh\left(1/\theta\right)\, $ with $\,\theta = 2 k_B T/\hbar \omega_0$. This is a very interesting result. At low temperature the interacting potential agrees with that obtained from Lifshitz formula \cite{Lifshitz}:
\begin{equation}
\label{VLif}
V_{z>\lambda_T}\left(z,T\right)\simeq V_{Lif}\left(z,T\right)=-k_B T \alpha_0/4 z^3\,
\end{equation}
with very small corrections proportional to $e^{-k_0\lambda_T}$. As a consequence, there are three distance regimes given by (\ref{SL}), (\ref{LL}) and (\ref{V_L}), corresponding, respectively, to the conditions $z\ll 1/k_0$,
$1/k_0\ll z\ll \lambda_T$ and $1/k_0\ll\lambda_T\leq z$. We can say that London-van der Waals interaction $(\propto 1/z^3)$ changes to Casimir-Polder interaction $(\propto 1/z^4)$ and then Lifshitz asymptotic behavior $(\propto T/z^3)$ takes place as the distance between the atom and the wall increases.

For very high temperature, $\lambda_T\ll 1/k_0$, however, the Casimir-Polder regime disappears, since condition $1/k_0\ll z\ll \lambda_T$ can not be satisfied anymore; equation (\ref{V_L}) is then applicable to all relevant distance regimes ($z>\lambda_T$): for $k_B T\gg \hbar \omega_0$, $f\left(\theta\right)\rightarrow 1$ and hence
$V\left(z,T\right)\simeq  -\hbar \omega_0 \alpha_0/8 z^3 =: V_{L}\left(z\right)$ and the London-van der Waals behavior dominates over all distances. We have not considere the regime where $z < \lambda_T \ll 1/k_0$ since it is not relevant experimentally.

\begin{figure}[!h]%
\begin{center}%
\vskip -0.8 cm%
\includegraphics[width=3.3in]{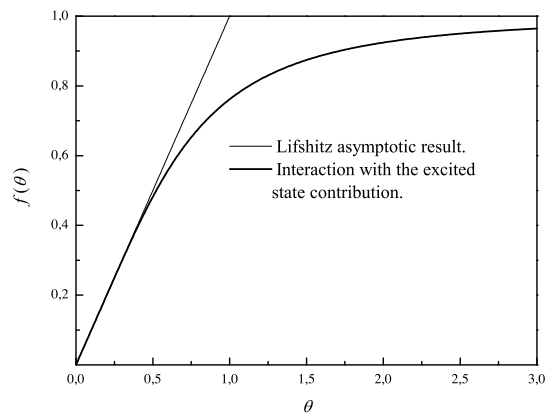}%
\vskip -0.8 cm%
%
\caption{Ratio  $V\left(z,T\right)/V_{L}\left(z\right)=f\left(\theta\right)$, where $V_L(z)=-\hbar\omega_0\alpha_0/8z^3$, in terms of $\theta = 2 k_B T/\hbar \omega_0$. The behaviour of ratio $V_{Lif}\left(z,T\right)/V_{L}\left(z\right)=\theta$ is also shown in the figure, for comparison.}
\label{VThigh}
\end{center}%
\end{figure}

\begin{figure}[!h]%
\begin{center}%
\vskip -1.0 cm%
\includegraphics[width=3.3in]{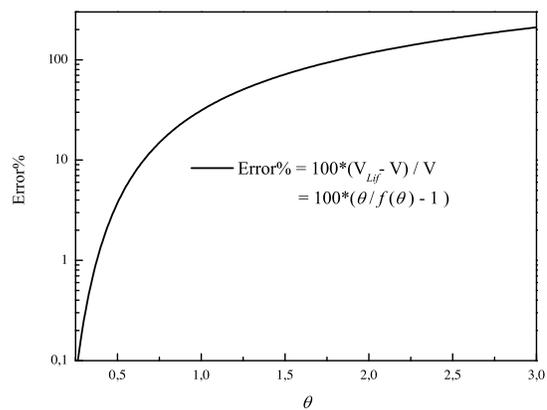}%
\vskip -0.8 cm%
%
\caption{Percentual error between the potential $V\left(z,T\right)$ given by (\ref{V_L}) and Lifshitz asymptotic result $V_{Lif}=-k_B T\alpha_0/4z^3$ in terms of $\theta = 2 k_B T/\hbar \omega_0$.}
%
%
\label{Error}%
\end{center}%
\end{figure}

\begin{figure}[!h]%
\begin{center}%
\vskip -0.8 cm%
\includegraphics[width=3.0in]{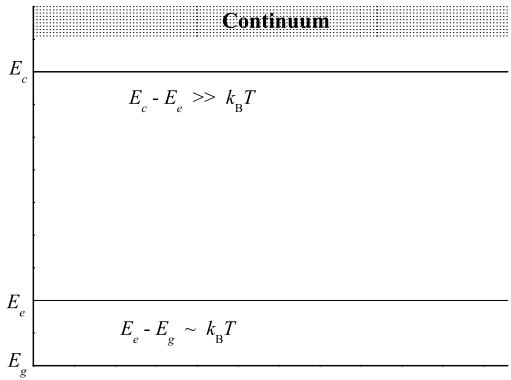}%
\vskip -0.5 cm%
%
\caption{A substance which simulates a two-level system that can be used to check our result (\ref{V_L}).  Levels $E_e$ and $E_g$ are sufficiently close to each other so that thermal fluctuations can populate significantly both of them: $E_e-E_g \sim k_B T\simeq 1/40\; \textrm{eV}$. Level $E_c$ is so far $E_e$ and $E_g$  ($E_c-E_e \gg k_B T$)
so that, in pratice, it can not be populated by thermal fluctuations.}
%
\label{gap}%
\end{center}%
\end{figure}

 In Figure \ref{VThigh} we plot  $V_{z>\lambda_T}(z,T)$ and $V_{Lif}(z,T)$ given by  (\ref{V_L}) and (\ref{VLif}) in terms of $\theta = 2k_B T/\hbar \omega_0$.  In the Figure \ref{Error} we plot the relative discrepancy $\Delta V\% = 100 \times \vert 1- V_{Lif}/V\vert$, with $V$ given by (\ref{V_L}).  One can see that for $\theta\simeq 0.26$, the discrepancy is $\Delta V\%\simeq 0.1\%$, while for $\theta\simeq 3.0$ this discrepancy raises to $\Delta V\%\simeq 210\%$. These  values would be easily detected. Hence, it is possible to check the results shown in Figure  \ref{Error} and the validity of equation (\ref{V_L}), at room temperature, using slabs of substances with energy level structures like the one shown in Figure \ref{gap}. Note that the energy level profile shown in this figure has two levels $E_g$ and $E_e$ sufficiently
 near to each other so that $E_e-E_g \sim k_B T$ and  another one $E_c$ such that $E_c-E_e \gg k_B T$.
%
%
%
%

 In order to calculate the force between a semi-infinite dilute slab of any material and a very good conductor, one may integrate  equation (\ref{V_L}) over all distances between the molecules of the slab and the conducting wall. Then, the dependence on $z$ and $T$ will be the same as given by (\ref{V_L}), though the constant of proportionality may be different due to the non-additivity of van der Waals forces \cite{PowerThiru}. Considering a distance $a$ between the conductor and a slab caracterized by a dielectric constant $\epsilon\sim 1$, one may write for the force per unit area, after performing the integration of (\ref{V_L}):%
\begin{equation}
F\left(z,T\right) \simeq -{3\hbar\omega_0\over 32\pi}
{f\left(\theta\right)\over a^3}\left({\epsilon-1\over \epsilon+2}\right)\, ,
\end{equation}
where we used the Clausius-Mosotti relation \cite{Jackson}: $4\pi \alpha_0 N = 3\left(\epsilon-1\right)/\left(\epsilon+2\right)$ with $N$ being the number of particles per unit volume of the rarefied medium with dielectric constant $\epsilon$. Since for this case the force $F\left(z,T\right)$  has the same behavior with tempearture as that given by (\ref{V_L}), the quantity $\Delta F\%=100\times\vert 1 - F_{Lif}\left(z,T\right)/F\left(z,T\right)\vert$, where  $F_{Lif}\left(z,T\right)\simeq -3k_B T\left(\epsilon-1\right)/16\pi\left(\epsilon+2\right) z^3$ is the force between the slabs
derived from $V_{Lif}\left(z,T\right)$, will behave exactly as shown in Figure (\ref{Error}).

In this letter, we have analyzed the contribution of the excited state of a two-level atom to the interaction potential
between  the atom and a perfectly conducting wall. We have shown that the corrections to London-van der Waals, Casimir-Polder and Lifshitz limits at low temperature ($k_0\lambda_T \gg 1$) are very small. For $k_0\lambda_T \sim 1$ or higher the interaction may differ strongly from Lifshitz result for distances $z\sim \lambda_T$, as shown in Figure \ref{Error}. Finally, for very high temperature ($k_0\lambda_T \ll 1$) the interaction behaves as the London-van der Waals limit for all distances. We expect that setups like those usually employed to measure  Casimir forces \cite{Bressi,Mohideen} may be used to check our results.
\vskip 0.1 cm

\noindent {\bf Acknowledgements}

\vskip 0.1 cm
\noindent TNCM and CF would like to thank Capes and CNPq, resepectively, for partial financial support.
\footnotesize

{

\end{document}